\documentclass{mem}
\usepackage{natbib}\usepackage{txfonts}\usepackage{balance}
\usepackage{graphicx}
\usepackage[a4paper]{hyperref}
\idline{75}{282}
\begin{document}
\def\teff{$T\rm_{eff }$}
\def\kms{$\mathrm {km s}^{-1}$}

\title{
Radio spectra of intermediate-luminosity broad-line radio galaxies
}

   \subtitle{}

\author{
E. \,Angelakis\inst{1}, M. \,Kadler\inst{2,3,4}, K. \,Lewis\inst{5}, R.\,M. \,Sambruna\inst{6}, M. \,Eracleous\inst{7},     
\and J.\,A.~\,~Zensus\inst{1}}

  \offprints{E. Angelakis}

\institute{
Max-Plank-Institut f\"ur Radioastronomie, Auf dem H\"ugel 69, DE-53121 Bonn, Germany
\email{angelaki@mpifr-bonn.mpg.de}
\and
Dr. Karl Remeis--Observatory,University of Erlangen-Nuremberg, Sternwartstrasse 7, Bamberg, 96049, Germany
\and
CRESST/NASA Goddard Space Flight Center, 662 Greenbelt, MD 20721, USA
\and
Universities Space Research Association, 10211 Wincopin Circle, Suite 500 Columbia, MD 21044, USA
\and
Department of Physics \& Astronomy, Dickinson College, Carlisle, PA 17013, USA
\and
Astroparticle Physics Laboratory, NASA Goddard Space Flight Center, Greenbelt, MD 20771, USA
\and
Department of Astronomy \& Astrophysics, The Pennsylvania State University, University Park, PA 16802, USA
}

\authorrunning{Angelakis }

\titlerunning{Radio spectra of intermediate-luminosity BLRGs}

\abstract{Within the context of investigating possible differences
  between the mechanisms at play in Radio Loud AGN and those in Radio
  Quiet ones, we study the spectral characteristics of a selected
  sample of Intermediate-Luminosity Broad-Line Radio Galaxies in
  X-rays, optical, IR and radio. Here, we present the radio spectra
  acquired with the 100-m radio telescope in Effelsberg between 2.6
  and 32 GHz. These measurements reveal a large variety of spectral
  shapes urging for radio imaging that would disclose the source
  morphology. Such studies could potentially discriminate between
  different mechanisms.   }
\maketitle{}

\section{Introduction}

A great deal of our current understanding of the processes at play in
AGNs is coming from studies of chiefly Radio-Quiet (RQ) objects. In
fact, it appears that different mechanisms become important at
different luminosities so that the emission from luminous RQ systems
is dominated by accretion flow \citep{Haardt1991ApJ}, whereas in lower
luminosity systems the accretion flow may be radiatively less
efficient \citep{Ho1999ApJ}.

On the other hand, little is known about the mechanisms determining
the behavior of the Radio-Loud (RL) AGNs, mostly because they comprise
only a small fraction of the complete AGN population. Studies of the
Broad-Line Radio Galaxy (BLRGs) sub-class have been focusing on the
brightest and most luminous objects. This rough picture leaves several
questions unanswered such as whether there are differences between the
physics of RL and RQ systems. A uniform coverage of the parameter space
of the studied AGNs could shed light on many of these issues. In
particular, the detailed comparison of RL and RQ AGNs of similar
luminosity can provide significant insight.

To make up for the bias towards the most luminous sources we have
selected a sample of 6 Intermediate-Luminosity BLRGs (IL-BLRGs) and
studied their radio spectrum. The ultimate goal of this attempt would
be to determine their Spectral Energy Distribution from radio to
X-rays utilizing the {\sl XMM-Newton} observations that have been
performed for these sources (see Sambruna et al. in prep.). That would
potentially reveal the relative importance of the various emission
mechanisms. Further, radio observations in conjunction with X-ray
measurements can provide information about the relative contribution
of the disk and the jet and how that depends on luminosity. Here, we
present the results of the radio study only.
\section{The sample}
\begin{table}
\caption{The used receivers characteristics}
\label{tbl:rx}
\begin{center}
\begin{tabular}{ccccc}
\hline
\\
  Frequency &$T_{sys}$ &  Sensitivity  &  FWHM\\
  (GHz)     &  (K) &  (K/Jy)  &  (arcsec)\\
\hline
\\
2.64   &17       &1.5   &260\\
4.85   &27       &1.5   &146\\
8.40   &22       &1.3   &81\\ 
10.45  &$\sim$50 &1.3   &66\\ 
14.60  &$\sim$50 &1.1   &50\\ 
32.00  &77       &0.5  &25 
\\
\hline
\end{tabular}
\end{center}
\end{table}
In its full version, the project currently discussed, focuses on the
study of 11 IL-BLRGs with $L_{2-10\mathrm{kev}}$ of the order of
$10^{42}-10^{43}$ erg~s$^{-1}$. These objects are all Radio Loud and
have flux densities $F_{1.4 GHz}\ge 4$ mJy and lie at $z\le 0.1$. From
cross-correlating the main galaxy catalog from the SDSS Data Release 2
\citep{Abazajian2004AJ} with the NVSS \citep{Condon1998AJ} and FIRST
catalogs \citep{Becker1995ApJ,White1997ApJ}, \cite{Best2005MNRAS}
compiled a catalog of 2215 RL objects. From a sub-set of 429 sources
with $z\le 0.1$, \cite{Hao2005AJ-I} found 11 that show broad H$\alpha$
lines ($\mathrm{FWHM(H}\alpha)~>~1200$~km~s$^{-1}$). Those comprise
the sample under investigation. Here we present the radio data for six
of these sources all of which have been observed by both Effelsberg
and {\sl XMM-Newton} (to be presented in Sambruna et al., in prep.).

\section{Observations and data reduction}
The radio spectra reported here have been measured with the 100-m
telescope in Effelsberg quasi-simultaneously on March 28, 2007.
\subsection{Observations}
The observations were performed with the heterodyne receivers at 2.64,
4.85, 8.35, 10.45, 14.60 and 32.00 GHz, mounted on the secondary focus
(see Table~\ref{tbl:rx}). The receivers at 4.85, 10.45 GHz are
equipped with multiple feeds allowing differential observations.
The 32-GHz system is a correlation receiver that is also significantly
insensitive to linear atmospheric effects, whereas the others are all
single-feed systems.
The measurements were conducted with the newly installed adaptive
secondary reflector characterized by low surface RMS that induces
higher sensitivity (up to 50\% increase at 43 GHz). Nevertheless, the
limiting factor has been the atmosphere itself.
The so-called "cross-scan" observing technique has been applied. Its
essence lies in measuring the response of the antenna as it is 
slewed over the source position in both azimuthal and elevation
direction.
The advantage of this method is mainly the fact that it allows the
direct detection of cases of confusion and also the correction of
pointing offset errors.
The individual spectra have been measured quasi-simultaneously within
1 hr to guarantee that they are free of source variability of
time-scales longer than that.
\begin{table*}
  \caption{The flux densities measured at Effelsberg along with the NVSS entries. Note that the upper limits are given in terms of three times the rms noise. The SNR is $\ge 3$.}
\label{tbl:fluxes}
\begin{center}
\begin{tabular}{cr@{$\pm$}lr@{$\pm$}lr@{$\pm$}lr@{$\pm$}lr@{$\pm$}lr@{$\pm$}lr@{$\pm$}l}
\hline
\\
Source &\multicolumn{2}{c}{$S_{1.4}^{\dagger}$} &\multicolumn{2}{c}{$S_{2.64}$} &\multicolumn{2}{c}{$S_{4.85}$} &\multicolumn{2}{c}{$S_{8.35}$} &\multicolumn{2}{c}{$S_{10.45}$} &\multicolumn{2}{c}{$S_{14.6}$} &\multicolumn{2}{c}{$S_{32.00}$}\\
(NVSS J) &\multicolumn{2}{c}{(mJy)} &\multicolumn{2}{c}{(mJy)} &\multicolumn{2}{c}{(mJy)} &\multicolumn{2}{c}{(mJy)} &\multicolumn{2}{c}{(mJy)} &\multicolumn{2}{c}{(mJy)}     &\multicolumn{2}{c}{(mJy)} \\
\hline
\\
023140$-$011005 &11.6&0.6 &33.6&0.6 &39.8&2 &\multicolumn{2}{c}{$-$} &37.1&1.4  &\multicolumn{2}{c}{$-$} &41.8&2.0$^{\ddagger}$  \\
081040$+$481230 &38.8&1.5 &31.3&0.4 &20.7&0.6 &12.6&0.9 &14.9&2.7 &\multicolumn{2}{c}{$-$} &17.4&3 \\
090624$+$005758 &5.8&0.4 &\multicolumn{2}{c}{$-$} & 8.5&0.6 &12.1&2.1 &14.0&0.8 &16.9&2.1 &\multicolumn{2}{c}{$-$} \\
101806$+$000539 &11.5&1.5 &8.0&0.5  &5.5&0.6  & 4.2&0.3 &\multicolumn{2}{l}{$\le$3$\cdot$42.3} &3.6&0.9 &\multicolumn{2}{c}{$-$} \\
125027$+$001345 &54.3&1.7 &58.1&1.6 &52.6&1.3 &57.4&0.9 &61.5&2.1 &61.7&2.8 &62.3&0.4 \\
142041$+$015931 &11.7&0.6 &9.7&1.9 &8.1&0.6 &4.8&0.4 &4.2&0.7 &\multicolumn{2}{c}{$-$} &\multicolumn{2}{c}{$-$}
\\
\hline
\multicolumn{8}{l}{$^{\dagger}$extracted from the NVSS \citep{Condon1998AJ}}\\
\multicolumn{8}{l}{$^{\ddagger}$observed in July 2007}
\end{tabular}
\end{center}
\end{table*}
Let us underline that the beam is large enough to leave the source
structure unresolved. This is especially important when the
interpretation the radio spectra in terms of source morphology is
attempted as we discuss in Section \ref{sec:spectra}.

\subsection{Data Reduction}
Considerable effort has been put in applying some necessary
post-observation corrections to the raw data: (a) {\sl Pointing offset
  correction}, which is meant to account for the power loss caused by
the offset between the telescope position and the true source
position. (b) {\sl Gain correction}; The 100-m telescope is designed
according to the homology principle that preserves the parabolic shape
of the reflector. However, small scale deformations cause an
elevation-dependent change of its gain that could also cause power
loss. The gain correction accounts for this effect. (c) {\sl Opacity
  correction}, which corrects for the atmospheric opacity effect. (d)
Finally, the data are subjected to {\sl sensitivity correction}, which
means translating the source antenna temperature into absolute
flux density units. It is done by observing standard sources referred
to as calibrators \citep[][ and Kraus priv. com.]{Baars1977AnA,Ott1994AnA}.

\section{The spectra}
\label{sec:spectra}
\begin{figure*}[t!]
\resizebox{\vsize}{!}{\includegraphics[clip=true]{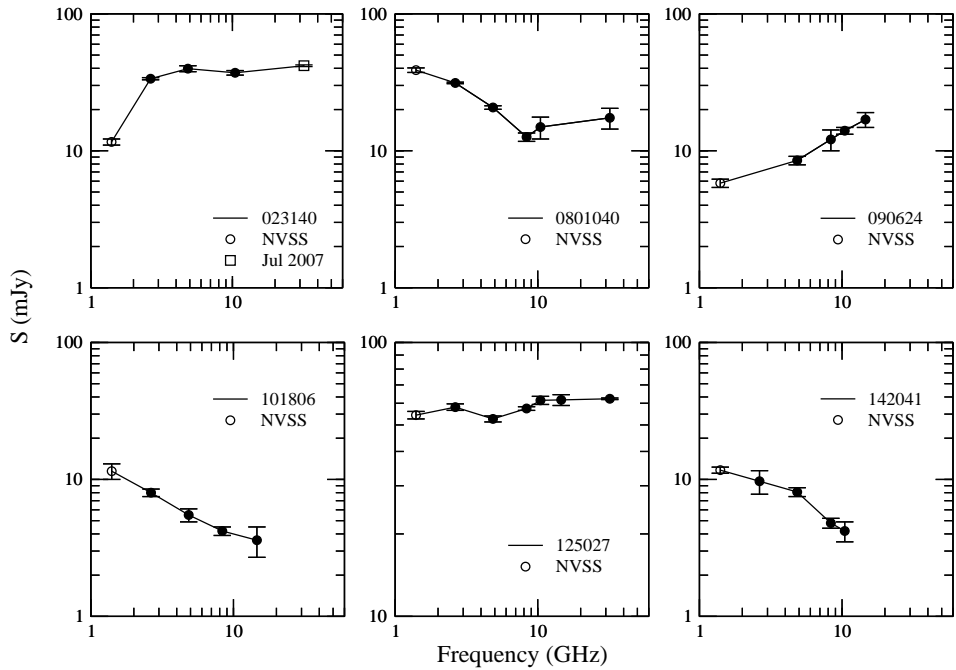}}
\caption{\footnotesize The measured spectra.}
\label{fig:spectra}
\end{figure*}

The acquired flux densities along with those from the NVSS catalog
\citep{Condon1998AJ} are summarized in Table~\ref{tbl:fluxes}. The
spectra are shown in Fig.~\ref{fig:spectra}. Only data with
$\mathrm{SNR}\ge 3$ have been used. Note that in
Table~\ref{tbl:fluxes} the upper limits are given in terms of three
times the RMS noise in the scan. It is noteworthy that the majority of
the targets are very weak making their measurements exceptionally
challenging.


From Fig. \ref{fig:spectra} it becomes clear that the sources possess
a large variety of spectral shapes. Henceforth, the name prefix ``NVSS
J'' is omitted from all the source names. 101806$+$000539 and
142041$+$015931 show a steep spectrum indicative of synchrotron jet
emission. 081040$+$481230 behaves similarly below 10 GHz (with $S\sim
{\nu}^{\alpha}$, $\alpha \approx-0.42$). At higher frequencies
however, it seems that a flat-spectrum component (possibly a core
component) becomes significant. Further observations are necessary to
test this idea. 125027$+$001345 is also a typical case of a flat
spectrum source ($\alpha \sim 0.05$) likely core-dominated. On the
other hand, 023140$-$011005 and 090624$+$005758 appear to be less
common. The former seems to be of flat spectrum with a low frequency
turnover that could also be caused by source variability (NVSS and
Effelsberg measurements are not simultaneous). The latter shows a
monotonically increasing spectrum (indicative of a GHz-peaked spectrum
source with $\alpha \sim 0.62$) suggesting a self absorbed system in
which we observe only the optically thick part of the
spectrum. Alternatively, such a spectral shape could be attributed to
a flat spectrum source (core-dominated) that undergoes a flaring
event. This could only be clarified with additional observations.

However, our recent re-analysis of the optical spectra showed that two
of the sources discussed earlier, namely 090624$+$005758 and
125027$+$001345, display optical spectral characteristics that could
classify them as NLRGs rather than as BLRGs. Having this solidly
confirmed would alter the interpretation of the spectra of those two
objects and necessitate the separate study of the two classes.

\section{Conclusions}
Clearly, the 6 BLRGs discussed here show a variety of radio spectra that
could be interpreted in different ways. Interferometric observations
would be essential for the correct interpretation of the observed
phenomenology. All these results in conjunction with the X-ray and
optical data are in preparation for publication (Sambruna et al., in
prep.).
\begin{acknowledgements}
  Based on observations with the 100-m telescope of the MPIfR
  (Max-Planck-Institut f\"ur Radioastronomie). MK and KL have been
  supported by the NASA Postdoctoral Program at the Goddard Space
  Flight Center, administered by Oak Ridge Associated Universities
  through a contract with NASA.
\end{acknowledgements}

\bibliographystyle{aa}

\end{document}